\documentstyle[12pt]{article}

\begin{document}
\title{Magnetic fingerprints on the spectra of one-electron and two-electrons
 interacting in parabolic quantum dots}
\author{ Omar Mustafa and Maen Odeh\\
 Department of Physics, Eastern Mediterranean University\\
 G. Magusa, North Cyprus, Mersin 10 - Turkey\\
 email: omar.mustafa@emu.edu.tr\\
\date{}\\}
\maketitle
\begin{abstract}
{\small Magnetic fingerprints on the spectra of an electron interacting
with a negatively charged ion in a parabolic quantum dot (QD), and of
two interacting electrons
in such a dot, are investigated via a new pseudoperturbative methodical
proposal. The effects of ion - electron and electron - electron interactions
on the spectra are studied. The effect of the symmetry of such problem is
emphasized. Compared with those obtained by Zhu et al.[6], via a series
solution, the results are found in excellent accord. Higher excited -
states are also reported.}
\end{abstract}
\newpage

\section{Introduction}
Quantum dots (QDs) are essentially quasi - zero - dimensional little islands,
populated by two - dimensional ( {\em flatland}) electrons laterally confined
by an artificial potential [1,2]. The variation of their spectra with
magnetic field is often called magnetic fingerprints of QDs.
Recent intensive experimental and
theoretical investigations have been carried out to explore various
spectral properties of QDs. Yet such studies are in their infancy and
still expanding rapidly [1-15].

The ion - electron and electron - electron interactions in such QDs
are of great importance [6,8-11]. The Hamiltonians (section 2-1) of which
are known to belong to non - exactly soluble Hamiltonians. One has therefore
to resort to approximation methods to study such systems. 
There has been, in the last few years, an increasing interest
in the study of two - electron QDs in a magnetic field [6,12-15].
Among the several eligible and reliable methods for solving the multi -
electron problem in QDs exist: exact numerical diagonalization [2,16],
numerical simulations based on quantum Monte Carlo techniques [17],
Hartree - Fock calculations [16, 18-20], a series solution based on
asymptotically physical grounds of the wave functions in the regions
$0<r$ and $r<\infty$ [6,21], etc. It is
therefore interesting to carry out systematic studies of the electronic
structures in QDs populated by one and two electrons with and
without magnetic fields.

Recently, we have introduced a pseudoperturbative ( artificial in nature)
shifted - $l$ ( $l$ is the angular momentum quantum number) expansion
technique ( PSLET) to solve Schr\"odinger equation for
states with arbitrary number of nodal zeros. It simply consists of using
$1/\bar{l}$ as a pseudoperturbation parameter, where $\bar{l}=l-\beta$
and $\beta$ is a suitable shift to be determined [22-27].
Encouraged by its very satisfactory performance ( documented in [22-27]
through comparisons with direct numerical integration, quasi perturbative,
variational, Hill determinant, etc, methods), we use
PSLET recipe, in section 2-2, and study the spectral properties of an
electron in a parabolic QD with a negatively charged ion ( impurity),
and two electrons in such a QD with and without the magnetic field.

In section 3 we compare our results with the ones obtained by Zhu et al.[6],
via a series solution, and report on other states that have not been
considered before, to the best of our knowledge. The effect of ion - electron
( section 3-1) and electron - electron ( section 3-2) interactions are
also discussed. In the absence of these interactions we show that whilst
the magnetic field {\em completely} lifts the confinement degeneracy, as it
increases from zero, for the one electron case, it {\em partially} lifts
it for the two - electron case. Moreover, at infinite magnetic fields the
energy levels cluster around Landau ones, inducing in effect Landau
degeneracy. The inclusion of these interactions, on the other hand,
{\em completely} lifts the confinement and Landau degeneracies and change
levels ordering. Consequently, energy levels crossing occur and spin -
singlet spin - triplet oscillations are manifested for two - electron QDs
in a magnetic field. Section 4 is devoted for concluding remarks.

\section{Hamiltonian models and PSLET recipe}

\subsection{Hamiltonian models}

To a very good approximation, the harmonic oscillator describes the lateral
confinement of electrons in some experimentally created QDs. The motion in
the z - direction ( the growth direction) is always frozen out into the
lowest subband [6,28]. The confinement in this direction is assumed to be
stronger than that in the xy-plane, and the dots, in effect, can be treated
as two - dimensional thin discs [6,21]. Then the Hamiltonian of an electron
in such a parabolic QD with a negatively charged ion center is very well
simulated by\\
\begin{equation}
H_{i0}=- \nabla^2 + \frac{1}{4} \gamma_{d}^2 ~\rho^2 + \frac{2}{\rho},
\end{equation}\\
and that of two electrons in the same QD by\\
\begin{equation}
H_0 = -\nabla_1^2 -\nabla_2^2 + \frac{1}{4} ~\gamma_{d}^2 ~(\rho_1^2
+ \rho_2^2) + \frac{2}{|\vec{\rho_1}-\vec{\rho_2}|},
\end{equation}\\
( in effective atomic units) with $ \gamma_d^{-1/2}$ related to the
confinement region ( hence to the quantum size) of the electrons in
the QDs. The energy and length are
given in effective Rydberg $R^*$ and effective Bohr radius $a^*$,
respectively.

When a magnetic field $\vec{B}$ is applied perpendicular to the xy - plane,
through the symmetric gauge $\vec{A}=(-y,x,0)B/2$, the Hamiltonians
in (1) and (2), respectively, read\\
\begin{equation}
H_i=H_{i0}+ \frac{1}{4} \gamma^2 ~\rho^2 + \gamma L_z,
\end{equation}\\
and\\
\begin{equation}
H=H_0 + \frac{1}{4} \gamma^2 ~(\rho_1^2 + \rho_2^2)
+ \gamma L_{z1} + \gamma L_{z2}.
\end{equation}\\
Where $\gamma=\hbar\omega_c/2R^*$ with
the cyclotron frequency $ \omega_c$, and $\gamma L_{z1}$ and
$\gamma L_{z2}$ are the induced Zeeman terms. Obviously, Hamiltonian (4)
is separable and can be recast as \\
\begin{equation}
H = H_R + H_r,
\end{equation}\\
where\\
\begin{equation}
H_R = - \frac{\nabla_R^2}{2} + \frac{1}{2} \Gamma^2 ~ R^2 + \gamma ~ L_{ZR},
\end{equation}\\
and\\
\begin{equation}
H_r = - 2~\nabla_r^2 + \frac{1}{8} \Gamma^2 ~ r^2 + \gamma ~ L_{zr}
+\frac{2}{r},
\end{equation}\\
which represent the center - of - mass (CM) and the relative - motion (RM)
Hamiltonians, respectively. Here, $\Gamma^2 = \gamma^2 + \gamma_d^2$,
$\vec{R} = (\vec{\rho_1}+\vec{\rho_2})/2$, $\nabla_R= \nabla_1+\nabla_2$,
$\vec{r} = \vec{\rho_1}-\vec{\rho_2}$, $\nabla_r = (\nabla_1-\nabla_2)/2$,
and $L_{ZR}=-i\partial/\partial\phi$ and $L_{zr}=-i\partial/\partial\varphi$
are the Z-  and z- components of the angular momentum operators in the CM
and RM systems, respectively. Effectively, the problem is reduced into two:
a quasi - particle of the CM coordinate in a rescaled external field, and
a quasi - particle of the RM coordinate in a rescaled external field and a
rescaled repulsive Coulomb field (emerging from the electron - electron
interaction). The cylindrical
symmetry of the attendant problems invites the separability of the wave
functions to obtain. Hence, the two - particle wave function in the plane
polar coordinate systems is $\Phi_{K,M}(R) \Phi_{k,m}(r) e^{im\varphi}
e^{iM\phi}$. The Pauli principle demands that $\Phi_{k,m}(r) e^{im\varphi}$
is antisymmetric with respect to inversion $\vec{r}\rightarrow
-\vec{r}$. No restrictions on $\Phi_{K,M}(R) e^{iM\phi}$ are imposed. One
would therefore obtain spin - singlet ($s=0$) and spin - triplet ($s=1$)
states for even and odd $m$, respectively, through the prescription
$s=(1-(-1)^m)/2$. Eventually, Schr\"odinger equation for Hamiltonian (7)
reads\\
\begin{equation}
\left[ - \frac{d^2}{dr^2} + \frac{m^2-1/4}{r^2} +\frac{1}{r}
+\frac{1}{16}\Gamma^2 r^2 \right] U_{k,m}(r) =
\left[\frac{E(k,m)-m\gamma}{2}\right] U_{k,m}(r),
\end{equation}\\
where $U_{k,m}(r)= \sqrt{r} \Phi_{k,m}(r)$, and $k$ and $m$ are the
radial and azimuthal quantum numbers in the RM system, respectively.

It is convenient to define the electron - electron interaction energies as\\
\begin{equation}
E_{ee}(k,m) = E(k,m) - E_0(k,m),
\end{equation}\\
where\\
\begin{equation}
E_0(k,m) = (2~k + |m| +1) \Gamma + m \gamma
\end{equation}\\
are the energies of (8) without the Coulomb interaction. Then the total
energies of (4) are\\
\begin{equation}
E(k,m;K,M) =E(k,m)+E(K,M),
\end{equation}\\
where\\
\begin{equation}
E(K,M)= (2~K + |M| + 1)\Gamma +M \gamma
\end{equation}\\
are the energies of (6), with $K$ and $M$ the radial
and azimuthal quantum numbers in the CM system, respectively.

Similarly, the ion - electron interaction energies are defined as\\
\begin{equation}
E_{ie}(k,m)=E_i(k,m)-E_{i0}(k,m),
\end{equation}\\
where\\
\begin{equation}
E_{i0}(k,m)= (2~k + |m| + 1)~\Gamma +m \gamma
\end{equation}\\
are the energies of (3) without the ion - electron interaction and
$E_i(k,m)$ are the eigenenergies of $H_i$ in (3).

\subsection{PSLET recipe}

In the underlying radical radial time - independent Schr\"odinger equation,
in $\hbar=m=1$ units,\\
\begin{equation}
\left[-\frac{1}{2}\frac{d^{2}}{dq^{2}}+\frac{l(l+1)}{2q^{2}}+V(q)\right]
\Psi_{k,l}(q)=E_{k,l}\Psi_{k,l}(q),
\end{equation}\\
the isomorphism between orbital angular momentum $l$ and  dimensionality
$D$ invites interdimensional degeneracies to obtain. Which, in
effect, adds a delicate nature to this equation and
 allows us to generate the ladder of excited states for any given $k$
( number of the nodal zeros in the wave function)
and nonzero $l$ from the $l=0$ result, with that $k$, by the transcription
$D \longrightarrow D+2l$. For more details the reader may
refer to ref.s [26,29-31].

Our recipe starts with shifting the angular momentum quantum number $l$
in (15) through $\bar{l} = l - \beta$ and use $1/\bar{l}$ as a
pseudoperturbation expansion parameter. Hence, equation (15) reads\\
\begin{equation}
\left\{-\frac{1}{2}\frac{d^{2}}{dq^{2}}+\frac{\bar{l}^{2}+(2\beta+1)\bar{l}
+\beta(\beta+1)}{2q^{2}}+\frac{\bar{l}^2}{Q}V(q) \right\}
\Psi_{k,l} (q)=E_{k,l}\Psi_{k,l}(q),
\end{equation}\\
where Q is a constant that scales the potential $V(q)$ at large - $l_D$ limit
and is set, for any specific choice of
$l_D$ and $k$, equal to $\bar{l}^2$ at the end of the calculations. Here
$l_D=l+(D-3)/2$, to incorporate the interdimensional degeneracies. Hence,
$\bar{l} \longrightarrow \bar{l}=l_D - \beta$ and $D=2$ ( with $l=|m|$)
through out this paper. Next,
we shift the origin of the coordinate system through
$x=\bar{l}^{1/2}(q-q_{o})/q_{o}$, where $q_o$ is currently an arbitrary
point to be determined below. Expansions about this point, $x=0$ (i.e.
$q=q_o$) would lead to\\
\begin{equation}
\left[-\frac{1}{2}\frac{d^{2}}{dx^{2}}+\frac{q_{o}^{2}}{\bar{l}}
\tilde{V}(x(q))\right]
\Psi_{k,l}(x)=\frac{q_{o}^2}{\bar{l}}E_{k,l}\Psi_{k,l}(x),
\end{equation}\\
with\\
\begin{equation}
\frac{q_o^2}{\bar{l}}\tilde{V}(x(q))=
q_o^2\bar{l}\left[\frac{1}{2q_o^2}+\frac{V(q_o)}{Q}\right]
+\bar{l}^{1/2}B_1 x+\sum^{\infty}_{n=0} v^{(n)}(x) \bar{l}^{-n/2},
\end{equation}\\
where\\
\begin{equation}
v^{(0)}(x)=B_2 x^2 + \frac{2\beta+1}{2},
\end{equation}\\
\begin{equation}
v^{(1)}(x)=-(2\beta+1) x + B_3 x^3,
\end{equation}\\
\begin{eqnarray}
v^{(n)}(x)&=&B_{n+2}~ x^{n+2}+(-1)^n~ (2\beta+1)~ \frac{(n+1)}{2}~ x^n
\nonumber\\
&+&(-1)^{n}~ \frac{\beta(\beta+1)}{2}~ (n-1)~ x^{(n-2)}~~;~~n \geq 2,
\end{eqnarray}\\
\begin{equation}
B_n=(-1)^n \frac{(n+1)}{2}
+\left(\frac{d^n V(q_o)}{dq_o^n}\right)\frac{q_o^{n+2}}{n! Q}.
\end{equation}\\
It is then convenient to expand $E_{k,l}$
 as\\
\begin{equation}
E_{k,l}=\sum^{\infty}_{n=-2}E_{k,l}^{(n)}~\bar{l}^{-n}.
\end{equation}\\
Equation (17), along with (18)-(22), is evidently the one - dimensional
Schr\"odinger equation for a harmonic oscillator $\Omega^2 x^2/2$,
with $\Omega^2=2B_2$, and the remaining terms in Eq.(17) as infinite power
series perturbations to the harmonic oscillator. One would then imply that\\
\begin{equation}
E_{k,l}^{(-2)}=\frac{1}{2q_o^2}+\frac{V(q_o)}{Q}
\end{equation}\\
\begin{equation}
E_{k,l}^{(-1)}=\frac{1}{q_o^2}\left[\frac{2\beta+1}{2}
+(k +\frac{1}{2})\Omega\right]
\end{equation}\\
Where $q_o$ is chosen to minimize $E_{k,l}^{(-2)}$, i. e.\\
\begin{equation}
\frac{dE_{k,l}^{(-2)}}{dq_o}=0~~~~
and~~~~\frac{d^2 E_{k,l}^{(-2)}}{dq_o^2}>0.
\end{equation}\\
Equation (26) in turn gives, with $\bar{l}=\sqrt{Q}$,\\
\begin{equation}
l_D-\beta=\sqrt{q_{o}^{3}V^{'}(q_{o})}.
\end{equation}\\
The shifting parameter $\beta$ is determined by choosing
$\bar{l}E_{k,l}^{(-1)}$=0. Hence\\
\begin{equation}
\beta=-\left[\frac{1}{2}+(k+\frac{1}{2})\Omega\right],~~
\Omega=\sqrt{3+\frac{q_o V^{''}(q_o)}{V^{'}(q_o)}}
\end{equation}\\
where primes of $V(q_o)$ denote
derivatives with respect to $q_o$. Then equation (17) reduces to\\
\begin{equation}
\left[-\frac{1}{2}\frac{d^2}{dx^2} + \sum^{\infty}_{n=0} v^{(n)}
\bar{l}^{-n/2}\right]\Psi_{k,l} (x)= 
\left[\sum^{\infty}_{n=1} q_o^2 E_{k,l}^{(n-1)}
\bar{l}^{-n} \right] \Psi_{k,l}(x).
\end{equation}\\
Setting the wave functions with any number of nodes $k$ as \\
\begin{equation}
\Psi_{k,l}(x(q)) = F_{k,l}(x)~ exp(U_{k,l}(x)),
\end{equation}\\
equation (29) readily transforms into the following Riccati equation:\\
\begin{eqnarray}
&&F_{k,l}(x)\left[-\frac{1}{2}\left( U_{k,l}^{''}(x)+U_{k,l}^{'}(x)
U_{k,l}^{'}(x)\right)
+\sum^{\infty}_{n=0} v^{(n)}(x) \bar{l}^{-n/2} \right. \nonumber\\
&&\left. -\sum^{\infty}_{n=1} q_o^2 E_{k,l}^{(n-1)} \bar{l}^{-n} \right]
-F_{k,l}^{'}(x)U_{k,l}^{'}(x)-\frac{1}{2}F_{k,l}^{''}(x)=0,
\end{eqnarray}\\
where the primes denote derivatives with respect to $x$. It is
evident that this equation admits solution of the form \\
\begin{equation}
U_{k,l}^{'}(x)=\sum^{\infty}_{n=0} U_{k}^{(n)}(x)~~\bar{l}^{-n/2}
+\sum^{\infty}_{n=0} G_{k}^{(n)}(x)~~\bar{l}^{-(n+1)/2},
\end{equation}\\
\begin{equation}
F_{k,l}(x)=x^k +\sum^{\infty}_{n=0}\sum^{k-1}_{p=0}
a_{p,k}^{(n)}~~x^p~~\bar{l}^{-n/2},
\end{equation}\\
where\\
\begin{equation}
U_{k}^{(n)}(x)=\sum^{n+1}_{m=0} D_{m,n,k}~~x^{2m-1} ~~~~;~~~D_{0,n,k}=0,
\end{equation}\\
\begin{equation}
G_{k}^{(n)}(x)=\sum^{n+1}_{m=0} C_{m,n,k}~~x^{2m}.
\end{equation}\\
Substituting equations (32) - (35) into equation (29) implies\\
\begin{eqnarray}
&&F_{k,l}(x)\left[-\frac{1}{2}\sum^{\infty}_{n=0}\left(U_{k}^{(n)^{'}}
\bar{l}^{-n/2}
+ G_{k}^{(n)^{'}} \bar{l}^{-(n+1)/2}\right) \right. \nonumber\\
&-&\left.\frac{1}{2} \sum^{\infty}_{n=0} \sum^{n}_{m=0}
\left( U_{k}^{(m)}U_{k}^{(n-m)} \bar{l}^{-n/2}
+G_{k}^{(m)}G_{k}^{(n-m)} \bar{l}^{-(n+2)/2}
\right. \right.\nonumber\\
&+&\left.\left.2 U_{k}^{(m)}G_{k}^{(n-m)} \bar{l}^{-(n+1)/2}\right)
+\sum^{\infty}_{n=0}v^{(n)} \bar{l}^{-n/2}
-\sum^{\infty}_{n=1} q_o^2 E_{k,l}^{(n-1)} \bar{l}^{-n}\right] \nonumber\\
&-&F_{k,l}^{'}(x)\left[\sum^{\infty}_{n=0}\left(U_{k}^{(n)}\bar{l}^{-n/2}
+ G_{k}^{(n)} \bar{l}^{-(n+1)/2}\right)\right]-\frac{1}{2}F_{k,l}^{''}(x)
=0
\end{eqnarray}\\
The solution of equation (36) follows from the uniqueness
of power series representation. Therefore, for a given $k$ we equate the
coefficients of the same powers of $\bar{l}$ and $x$, respectively. 
One can then calculate
the energy eigenvalues and eigenfunctions from the knowledge of
$C_{m,n,k}$, $D_{m,n,k}$, and $a_{p,k}^{(n)}$ in a hierarchical manner.
Nevertheless, the procedure just described is suitable for a
software package such as  MAPLE to determine
the energy eigenvalue and eigenfunction corrections up to any order of the
pseudoperturbation series (23). 

Although the energy series, equation (23), could appear
divergent, or, at best, asymptotic for small $\bar{l}$, one can still 
calculate the eigenenergies to a very good accuracy by forming the 
sophisticated Pad\'e approximation [22-24,26,32]\\
\begin{equation}
P_{N}^{M}(1/\bar{l})=(P_0+P_1/\bar{l}+\cdots+P_M/\bar{l}^M)/
(1+q_1/\bar{l}+\cdots+q_N/\bar{l}^N)
\end{equation}\\
to the energy series (23). The energy series is calculated up to
$E_{k,l}^{(18)}/\bar{l}^{18}$ by
\begin{equation}
E_{k,l}=\bar{l}^{2}E_{k,l}^{(-2)}+E_{k,l}^{(0)}+\cdots
+E_{k,l}^{(18)}/\bar{l}^{18}+O(1/\bar{l}^{19}),
\end{equation}\\
and with the $P_{10}^{9}(1/\bar{l})$ Pad\'e approximant it becomes\\
\begin{equation}
E_{k,l}[10,9]=\bar{l}^{2}E_{k,l}^{(-2)}+P_{10}^{9}(1/\bar{l}).
\end{equation}\\
Our recipe is therefore well prescribed.

\section{Results and discussion}

It is obvious, to a scaling factor, that Hamiltonians (3) and (7) bear the
same form of a hybrid of a repulsive Coulomb and oscillator potentials\\
\begin{equation}
V(q)=b^2q^2/2+2/q.
\end{equation}\\
Hence, equation (28) yields\\
\begin{equation}
\Omega=\sqrt{\frac{4b^2~q_o^3-2}{b^2~q_o^3-2}},
\end{equation}\\
and,in turn, equation (27) reads\\
\begin{equation}
l_D+\frac{1}{2}\left(1+(2k+1)\sqrt{\frac{4~b^2~q_o^3-2}{b^2~q_o^3-2}} \right)
= \sqrt{b^2~q_o^4-2~q_o}.
\end{equation}\\
Once $q_o$ is determined (often numerically) the coefficients
$C_{m,n,k}$, $D_{m,n,k}$, and $a_{p,k}^{(n)}$ are obtained in a
sequential manner. Then the eigenvalues and eigenfunctions
are calculated in one batch for each value of $k$, $D$, $l$, and $b$.

In order to make remediable analysis of our results we have calculated the
first twenty terms of our energy series. We have also computed the
corresponding sequence of Pad\'{e} approximants $P^{1}_{2}(1/\bar{l})$,
$P^{2}_{2}(1/\bar{l})$,$\cdots$, $P^{9}_{9}(1/\bar{l})$,
$P^{9}_{10}(1/\bar{l})$ and observed their effects on the
leading energy term $\bar{l}^2 E_{k,l}^{(-2)}$. Moreover, the {\em twofold}
effect of the first term in the effective potential\\
\begin{equation}
V_{eff}(q)=\frac{m^2-1/4}{q^2}+V(q)
\end{equation}\\
should be in point. That is, for $|m|=0$ it represents an attractive core
that strengthens the comfinement, whereas for $|m|\ge1$ it represents a
repulsive core and strengthens the Coulomb repulsion.

\subsection{Ion - electron interaction effect}

In table 1 we display the energies of an electron in parabolic QDs, {\em
including} ion - electron interaction, in a magnetic field for $\gamma_d=0.2$.
They compare excellently with those reported by Zhu et al.
in figures 1(a) and 1(b) of [6]. In addition we report PSLET results
for 4d$^-$, 4p$^-$, 5d$^-$, and 6f$^-$ states.

In figure 1 we plot the
energies of an electron in such QDs ( with $\gamma_d=0.2$) versus $\gamma$,
{\em excluding} ion - electron interaction. Clearly, it shows that
whilst the magnetic field {\em completely} lifts the well known
confinement degeneracy, $(2k+|m|+1)\gamma_d$,
as $\gamma$ increases from zero, it eventually introduces Landau 
degeneracy as $\gamma\longrightarrow\infty$. And, at $\gamma=\infty$
only s - states are feasible and degenerate. That is, a state $(k^{'},m^{'})$
would cluster at an s-state $(k,0)$ through the prescription\\
\begin{equation}
2(k^{'}-k)=-|m^{'}|-m^{'}.
\end{equation}

In the range of small $\gamma$ ( namely, $0<\gamma<0.4$), one clearly
observes that there are minima for states with negative $m$, whereas
for states with positive $m$ and s-states
($m=0$) increase monotonically. This should be attributed to the following:
(i) for $m=0$, the effective potential does not support a minimum and the
Zeeman term has no effect on these states, hence s-states increase
monotonically. (ii) For $|m|\ge1$, the effective potential supports 
minima and the Zeeman term $m\gamma$ refines them for negative value of
$m$ or shifts them up for positive $m$. However, in the range of large
$\gamma$ the magnetic field ( in $\sim\Gamma^{2}q^{2}$) dominates over the
Zeeman term and the leading term in the effective potential, hence all energy
levels increase monotonically in this range of $\gamma$. As a result of (i)
and (ii) one observes the energy levels crossings between different states in
the range of small $\gamma$ in figure 1.

The inclusion of the ion-electron interaction ( figures 2-4) significantly
changes the spectral properties mentioned above. At $\gamma=0$ ( see
figures 2-4), one notices that all energies are shifted up. For example,
the 1s-states is shifted (in $Ry^{*}$ units) by 0.6162, 2p$^{-}$ and
2p$^{+}$ by 0.4666, 3d$^{-}$ and 3d$^{+}$ by 0.3829, 4f$^{-}$ and 4f$^{+}$
by  0.3305, etc. This shift decreases as $|m|$ increases. For higher levels
( larger $|m|$) $q$ increases and the Coulomb repulsion $\sim 1/q$ decreases,
a characteristic which is reflected upon the energy shift - ups in
figures 2-4. As $\gamma$ increases from zero, figures 2-4 clearly show
that the ion-electron interaction {\em completely} lifts Landau degeneracy
and results in change in levels ordering. Consequently, energy crossings
occur and odd-even parity oscillations are manifested ( 2p$^{+}$ crosses
with 4d$^{-}$, 4d$^{-}$ crosses with 4f$^{+}$, 3d$^{+}$ in figure 3).
Figure 1(b) of Zhu et al.[6] also bear this out.

To explain energy crossings, we consider the energy shift-ups as a result
of the ion-electron interaction. At $\gamma=0.4$ ( energy crossings occur
in the range $0<\gamma<0.4$) for example, the 1s-state is shifted ( in
Ry$^{*}$ units) by $\sim 1.03$, 2p$^{-}$ and 2p$^{+}$ by $\sim 0.73$,
3d$^{-}$ and 3d$^{+}$ by $\sim 0.59$, 4f$^{-}$ and 4f$^{+}$ by $\sim 0.50$,
 and so on. Obviously the ion-electron interaction does not distinguish
between positive and negative $m$ values. Moreover, the
shift-ups decrease as $|m|$ increase for a given $k$, or for a given $|m|$
as $k$ increases ( because of the Coulomb repulsion characteristic mentioned
above).This in fact explains why the energy ladder of the lower batch, say,
 of energy levels in figure 1 is turned up-side-down in figure 2. Similar
 scenarios could be also developed for figures 3 and 4.

 \subsection{ Electron - electron interaction effect}

Next, we calculate the e$^-$ - e$^-$ interaction energies $E_{ee}(n_r,m)$
for two electrons in QDs and compare them, in tables 2 and 3, with those
reported by Zhu et al.[6]. They are in almost exact accord. We also display
PSLET results for 4s, 4p, 5d, 4f, and 5g states. Figure 5 shows that
$E_{ee}(n_r,m)$ increases with $\gamma$ and the levels ordering is
$E_{ee}(0,0) > E_{ee}(1,0) > E_{ee}(2,0) > \cdots > E_{ee}(0,1) >
E_{ee}(1,1) > E_{ee}(2,1)> \cdots > E_{ee}(0,2) > E_{ee}(1,2) > \cdots$, etc.

Before we proceed any further let us study the quantum size effect on the
two-electron spectra.
Table 4 shows the quantum levels of two electrons in QDs with different
values of $\gamma_d(\gamma_d^{-1/2})$, hence showing the quantum size effect,
when $\gamma=0$. It documents changes in the levels ordering ( consequently,
energy crossings and singlet-triplet  spin  oscillations occur)
 as $\gamma_d(\gamma_d^{-1/2})$
decreases ( increases) from 1(1). Hereby, it should be noted that Zhu's [6]
result $E(1,0;0,1;0)$ for $\gamma_d=0.2$ is now corrected from 1,4402 to
1.4413 in accordance with his finding in table 1 of [6]. The levels orderings
for $\gamma_d=0.05$ are changed from (h) (e) to (e) (h) and from (k) (n) to
(n) (k) in accordance with our findings, based on the stability of, at least,
the last five terms of the Pad\'{e} sequence ( a signal of convergence to
the exact results) mentioned above.

To study the e$^{-}$-e$^{-}$ interaction effect on the spectra, we plot
the energy levels $E(k,m;K,M;s)$ of two-electron QDs with
$\gamma_d=0.2$ and at different values of $\gamma$ are plotted in figure 6
( {\em excluding} the e$^-$ - e$^-$ interaction) and figure 7 ( {\em
including} e$^-$ - e$^-$ interaction). Figure 6
clearly shows that the magnetic
field {\em partially} lifts the well known confinement degeneracy
 as $\gamma$ increases from zero and induces Landau
degeneracy as $\gamma\longrightarrow\infty$.
Again, one observes the clustering of the quantum levels around
$E(k,0;K,0;0)$ states as $\gamma\longrightarrow\infty$, following a
similar trend as that of (44).

The inclusion of the e$^-$ - e$^-$ interaction {\em completely} lifts the
confinement and Landau degeneracies and changes the levels ordering.
Consequently, energy crossings occur, at which singlet - triplet spin
oscillations obtain, at specific magnetic fields. Figures 6 and 7 bear these
out. One could follow a similar scenario as that in section 3-1 to explain
the energy crossings, since the e$^{-}$-e$^{-}$ interaction is effectively
a repulsive Coulomb term.

In table 5 we display our results for two interacting electrons in QDs in
a magnetic field at different values of $\gamma$, and for
$\gamma_d=0.2$. When plotted with the same scale, our results excellently
agree with those of Zhu, figure 2(b) in [6]. However, it should be
mentioned that the quantum levels classified in figure 2(b) of Zhu [6]
as a, b, c, $\cdots$ do not bear positive values of m and M but rather
negative ones, classified in table 5 of this text.

\section{Concluding remarks}

In this paper we have used our recently developed PSLET theory [22-27] to
study the magnetic fingerprints on the spectra of an electron in parabolic
QDs with negatively charged ion, and on two interacting electrons in such
QDs. We have also emphasized the effect of the symmetry of the problem
( marked in the leading term of the effective potential in (43)) on the
confinement and Coulomb repulsive terms. The comparison of PSLET results
with those of Zhu et al.[6] is readily very satisfactory.

Although we have started with the central force Schr\"odinger equation and
augmented the orbital angular momentum by $l \longrightarrow l_D=l+(D-3)/2$,
to incorporate interdimensional degeneracies, we have only considered the
$D=2$ with $l=|m|$ case for the attendant problems.

A general observation concerning the method used by Zhu et al.[6] is in
order. We have already mentioned that the series solution method used
by Zhu et al.[6] is based upon the asymptotically physical grounds of the wave
functions in the regions $0<r$ and $r<\infty$ ( i.e. $r\longrightarrow0$ and
$r\longrightarrow\infty$, respectively). Effectively and obviously, the
authors used the asymptotic behaviours of the wave functions at weak and
strong magnetic field limits implicitly. In the weak field limit
the wave function is dominated by a Coulombic character ( hence Coulombic
like basis are used in equation (12) of [6]) and in the strong field limit
its dominated by a harmonic oscillator character ( hence harmonic
oscillator like basis are used in equation (14) of [6]). In fact, this is
the only explanation, we could think of, as to why our results do not
exactly agree with those of Zhu in table 2, for $\Gamma=0.4$.

The conceptual soundness of our PSLET is obvious. It avoids troublesome
questions pertaining to the nature of small parameter expansions ( weak
or strong field limits), and trend of convergence to the exact results.
On the
computational and practical methodical sides, it offers ( beyond its promise
as being quite handy) a useful pseudoperturbation prescription where the
zeroth order approximation $\bar{l}^2~E^{(-2)}_{k,l}$ inherits a substantial
amount ( more than 90\% for the above problems) of the total energy.

Finally, PSLET theory could be applied to two electrons in an external
oscillator potential in $D=3$ space [33], QD lattices [34], 3 - electron
QDs [35], 2 - dimensional hydrogenic donor states in a magnetic field
[36-39], etc.

\newpage
\begin{table}
\begin{center}
\caption{ $E_i(k,m)$ with the ion - electron interaction for
$\gamma_d=0.2$ and different values of $\gamma$}
\vspace{1cm}
\begin{tabular}{|lccccc|}
\hline\hline
 $\gamma$ & 0 & 0.1 & 0.2 & 0.3 & 0.4\\
\hline
$E_i(0,0)$  (1s)     & 0.8162 & 0.8860 & 1.0537 & 1.2617 & 1.4816 \\
$E_i(0,-1)$ ($2p^-$) & 0.8666 & 0.8445 & 0.9339 & 1.0724 & 1.2282 \\
$E_i(0,-2)$ ($3d^-$) & 0.9829 & 0.8775 & 0.9100 & 1.0070 & 1.1307 \\
$E_i(0,-3)$ ($4f^-$) & 1.1305 & 0.9449 & 0.9278 & 0.9922 & 1.0922 \\
$E_i(0,1)$ ($2p^+$)  & 0.8666 & 1.0445 & 1.3339 & 1.6724 & 2.0288 \\
$E_i(0,2)$ ($3d^+$)  & 0.9829 & 1.2775 & 1.7100 & 2.2070 & 2.7307 \\
$E_i(0,3)$ ($4f^+$)  & 1.1305 & 1.5449 & 2.1278 & 2.7922 & 3.4922 \\
$E_i(1,0)$ ($2s$)    & 1.1741 & 1.2864 & 1.5612 & 1.9097 & 2.2867 \\
$E_i(1,-1)$ ($3p^-$) & 1.2336 & 1.2557 & 1.4563 & 1.7415 & 2.0623 \\
$E_i(1,-2)$ ($4d^-$) & 1.3599 & 1.2999 & 1.4466 & 1.6940 & 1.9860 \\
\hline\hline
$\gamma$ & 0 & 0.05 & 0.1 & 0.15 & 0.2 \\
\hline
$E_i(1,1)$ ($3p^+$)  & 1.2336 & 1.3156 & 1.4557 & 1.6405 & 1.8563 \\
$E_i(2,2)$ ($3s$)    & 1.5406 & 1.5815 & 1.6967 & 1.8692 & 2.0814 \\
$E_i(2,-1)$ ($4p^-$) & 1.6077 & 1.6014 & 1.6748 & 1.8101 & 1.9888 \\
$E_i(2,-2)$ ($5d^-$) & 1.7415 & 1.6900 & 1.7273 & 1.8339 & 1.9893 \\
$E_i(2,-3)$ ($6f^-$) & 1.9002 & 1.8042 & 1.8069 & 1.8869 & 2.0218 \\
\hline\hline
\end{tabular}
\end{center}
\end{table}
\newpage
\begin{table}
\begin{center}
\caption{ Electron - electron interaction energies $E_{ee}(k,m)$ at
different values of $\Gamma$. The values in brackets are reported by
Zhu et al.[6].}
\vspace{1cm}
\begin{tabular}{|ccccc|}
\hline\hline
$\Gamma$ & 0.05 & 0.1 & 0.2 & 0.4 \\
\hline
$E_{ee}(0,0)$ (1s) & 0.1963 & 0.3081 & 0.4816 & 0.7479 \\
                   & (0.1963) & (0.3081) & (0.4816) & (0.7494) \\
$E_{ee}(1,0)$ (2s) & 0.1853 & 0.2871 & 0.4413 & 0.6716 \\
                   & (0.1853) & (0.2871) & (0.4413) & (0.6723) \\
$E_{ee}(2,0)$ (3s) & 0.1763   & 0.2703 & 0.4106 & 0.6164  \\
                   & (0.1763) & (0.2703) & (0.4106) & (0.6000)  \\
$E_{ee}(3,0)$ (4s) & 0.1686   & 0.2565 & 0.3863 & 0.5742  \\
$E_{ee}(0,1)$ (2p) & 0.1562   & 0.2333 & 0.3451 & 0.5057  \\
                   & (0.1562) & (0.2333) & (0.3451) & (0.5066)  \\
$E_{ee}(1,1)$ (3p) & 0.1468   & 0.2168 & 0.3170 & 0.4597  \\
                   & (0.1468) & (0.2168) & (0.3170) & (0.4637) \\
$E_{ee}(2,1)$ (4p) & 0.1392   & 0.2039 & 0.2959 & 0.4265  \\
$E_{ee}(0,2)$ (3d) & 0.1311   & 0.1915 & 0.2776 & 0.4000  \\
                   & (0.1311) & (0.1915) & (0.2776) & (0.3998) \\
$E_{ee}(1,2)$ (4d) & 0.1240   & 0.1799 & 0.2594 & 0.3720  \\
                   & (0.1240) & (0.1799) & (0.2594) & (0.3784) \\
$E_{ee}(2,2)$ (5d) & 0.1182   & 0.1707 & 0.2452 & 0.3505  \\
$E_{ee}(0,3)$ (4f) & 0.1144   & 0.1652 & 0.2375 & 0.3399  \\
$E_{ee}(0,4)$ (5g) & 0.1025   & 0.1472 & 0.2105 & 0.3002  \\
\hline\hline
\end{tabular}
\end{center}
\end{table}

\newpage
\begin{table}
\begin{center}
\caption{ Same as table 2 for different values of $\Gamma$.}
\vspace{1cm}
\begin{tabular}{|ccccc|}
\hline\hline
$\Gamma$ & 1 & 2.5 & 4 & 5 \\
\hline
$E_{ee}(0,0)$ (1s) & 1.3196 & 2.2807 & 2.9934 & 3.3992 \\
                   & (1.3195) & (2.2803) & (2.9930) & (3.3988) \\
$E_{ee}(1,0)$ (2s) & 1.1473 & 1.9154 & 2.4721 & 2.7862 \\
                   & (1.1473) & (1.9154) & (2.4721) & (2.7862) \\
$E_{ee}(2,0)$ (3s) & 1.0333   & 1.6966 & 2.1744 & 2.4435  \\
                   & (1.0333) & (1.6967) & (2.1744) & (2.4435)  \\
$E_{ee}(3,0)$ (4s) & 0.9511   & 1.5469 & 1.9754 & 2.2166  \\
$E_{ee}(0,1)$ (2p) & 0.8279   & 1.3404 & 1.1707 & 1.9195  \\
                   & (0.8279) & (1.3404) & (1.7107) & (1.9195)  \\
$E_{ee}(1,1)$ (3p) & 0.7439   & 1.1941 & 1.5188 & 1.7017  \\
                   & (0.7438) & (1.1941) & (1.5188) & (1.7017) \\
$E_{ee}(2,1)$ (4p) & 0.6860   & 1.0964 & 1.3922 & 1.5589  \\
$E_{ee}(0,2)$ (3d) & 0.6436   & 1.0294 & 1.3077 & 1.4645  \\
                   & (0.6436) & (1.0294) & (1.3078) & (1.4645) \\
$E_{ee}(1,2)$ (4d) & 0.5957   & 0.9496 & 1.2047 & 1.3485  \\
                   & (0.5957) & (0.9496) & (1.2047) & (1.3485) \\
$E_{ee}(2,2)$ (5d) & 0.5598   & 0.8907 & 1.1292 & 1.2636  \\
$E_{ee}(0,3)$ (4f) & 0.5432   & 0.8650 & 1.0969 & 1.2277  \\
$E_{ee}(0,4)$ (5g) & 0.4782   & 0.7598 & 0.9628 & 1.0772  \\
\hline\hline
\end{tabular}
\end{center}
\end{table}
\newpage
\begin{table}
\begin{center}
\caption{ $E(k,m;K,M;s)$ quantum levels of two electrons in QDs at
different values of $\gamma_d(\gamma_d^{-1/2})$ for $\gamma=0$, including
electron - electron interaction. Zhu's results [6] are obtained by replacing
the last j digits of our results with the j digits in parentheses.}
\vspace{1cm}
\begin{tabular}{|lllll|}
\hline\hline
$\gamma_d (\gamma_d^{-1/2})$ & 1(1) & 0.4(1.5811) & 0.2(2.2361) & 0.05(4.4721) \\
\hline
a:(0,0;0,0;0) & a) 3.3196      & a) 1.5479 & a) 0.8816      & a) 0.2963(2) \\
b:(0,1;0,0;1) & b) 3.8279(8)   & b) 1.7057 & b) 0.9451(0)   & b) 0.3062 \\
c:(0,0;0,1;0) & c) 4.3196      & c) 1.9479 & d) 1.0776      & d) 0.3311(0) \\
d:(0,2;0,0;0) & d) 4.6436      & d) 2.0000 & c) 1.0816      & c) 0.3463(2) \\
e:(0,1;0,1;1) & e) 4.8279(8)   & e) 2.1057 & e) 1.1451(0)   & e) 0.3562 \\
f:(1,0;0,0;0) & f) 5.1473(2)   & f) 2.2716 & h) 1.2375(156)& h) 0.3644(476)\\
g:(0,0;1,0;0) & g) 5.3196      & h) 2.3399 & f) 1.2413(02)  & i) 0.3811(0) \\
h:(0,3;0,0;1) & h) 5.5432(174) & g) 2.3479 & i) 1.2776      & f) 0.3853(4) \\
i:(0,2;0,1;0) & i) 5.6436      & i) 2.4000 & g) 1.2816      & g) 0.3963(2) \\
j:(1,1;0,0;1) & j) 5.7439(8)   & j) 2.4597 & j) 1.3170      & j) 0.3968 \\
k:(0,1;1,0;1) & k) 5.8279(8)   & k) 2.5057 & k) 1.3451(0)   & n) 0.4025(66) \\
l:(1,0;0,1;0) & l) 6.1473(2)   & l) 2.6716 & n) 1.4105(053) & k) 0.4062 \\
m:(0,0;1,1;0) & m) 6.3196      & n) 2.7002 & l) 1.4413      & o) 0.4240 \\
n:(0,4;0,0;0) & n) 6.4782(93)  & m) 2.7479 & o) 1.4594      & p) 0.4311(0) \\
o:(1,2;0,0;0) & o) 6.5957(6)   & o) 2.7720 & p) 1.4776      & l) 0.4353(4) \\
p:(0,2;1,0;0) & p) 6.6436      & p) 2.8000 & m) 1.4816      & m) 0.4463(2) \\
\hline\hline
\end{tabular}
\end{center}
\end{table}
\newpage
\begin{table}
\begin{center}
\caption{ $E(k,m;K,M;s)$ quantum levels of two electrons in QDs with
different values of $\gamma$ and for $\gamma_d=0.2$, including electron -
electron interaction.}
\vspace{1cm}
\begin{tabular}{|ccccccc|}
\hline\hline
$\gamma$ & 0 & 0.05 & 0.1 & 0.2 & 0.3 & 0.4 \\
\hline
A:(0,0;0,0;0) & 0.8816 & 0.9033 & 0.9644 & 1.1664 & 1.4217 & 1.6967 \\
B:(0,-1;0,0;1) & 0.9451 & 0.9194 & 0.9380 & 1.0667 & 1.2599 & 1.4790 \\
C:(0,0;0,-1;0) & 1.0816 & 1.0595 & 1.0880 & 1.2493 & 1.4822 & 1.7439 \\
D:(0,-2;0,0;0) & 1.0776 & 1.0067 & 0.9889 & 1.0648 & 1.2211 & 1.4129 \\
E:(0,-1;0,-1;1) & 1.1451 & 1.0756 & 1.0616 & 1.1496 & 1.3201 & 1.5262 \\
F:(0,-3;0,0;0) & 1.2375 & 1.1220 & 1.0697 & 1.0984 & 1.2249 & 1.3960 \\
G:(1,0;0,0;0) & 1.2413 & 1.2742 & 1.3670 & 1.6766 & 2.0734 & 2.5066 \\
H:(1,-1;0,0;1) & 1.3170 & 1.3030 & 1.3548 & 1.5963 & 1.9378 & 2.3238 \\
I:(1,0;0,-1;0) & 1.4413 & 1.4304 & 1.4906 & 1.7594 & 2.1339 & 2.5538 \\
J:(1,-2;0,0;0) & 1.4594 & 1.4004 & 1.4166 & 1.6078 & 1.9158 & 2.2773 \\
K:(1,-1;0,-1;1) & 1.5170 & 1.4592 & 1.4784 & 1.6792 & 1.9984 & 2.3710 \\
\hline\hline
\end{tabular}
\end{center}
\end{table}
\newpage
\begin{center}
\bf{\Large Figures captions}
\end{center}

{\bf Fig.1:} $E_i(k,m)$ versus $\gamma$ for the ion - electron in QDs with
            $\gamma_d=0.2$, excluding the ion - electron interaction.\\ 

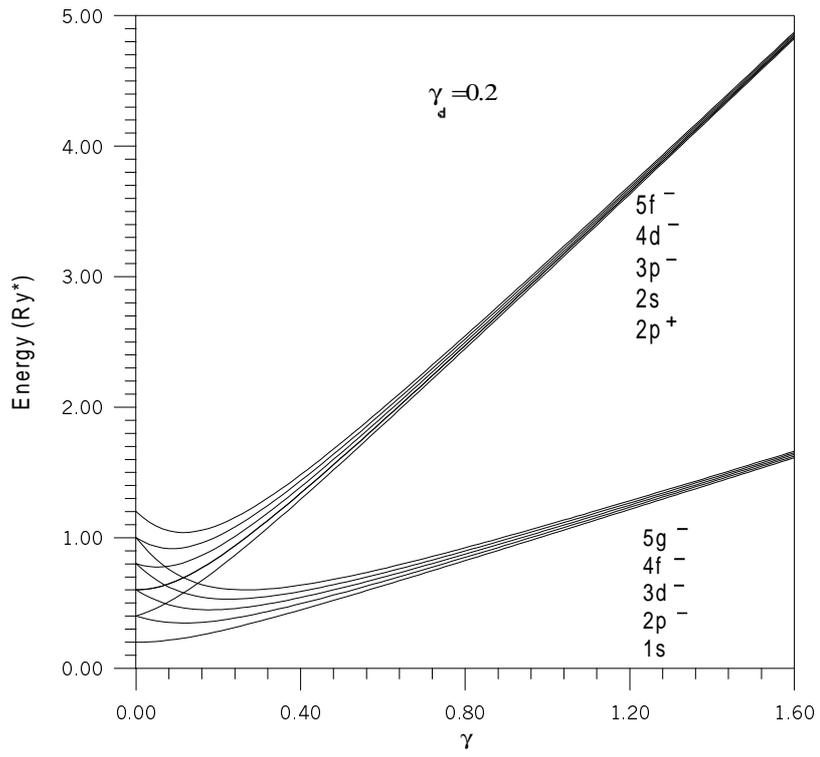
{\bf Fig.2:} Same as figure 1 including the ion - electron interaction for
            1s, 2p$^-$, 3d$^-$, and 4f$^-$ states.\\

{\bf Fig.3:} Same as figure 2 for 2s, 3d$^+$, 4f$^+$, 3p$^-$, 2p$^+$, and
            4d$^-$ states.\\

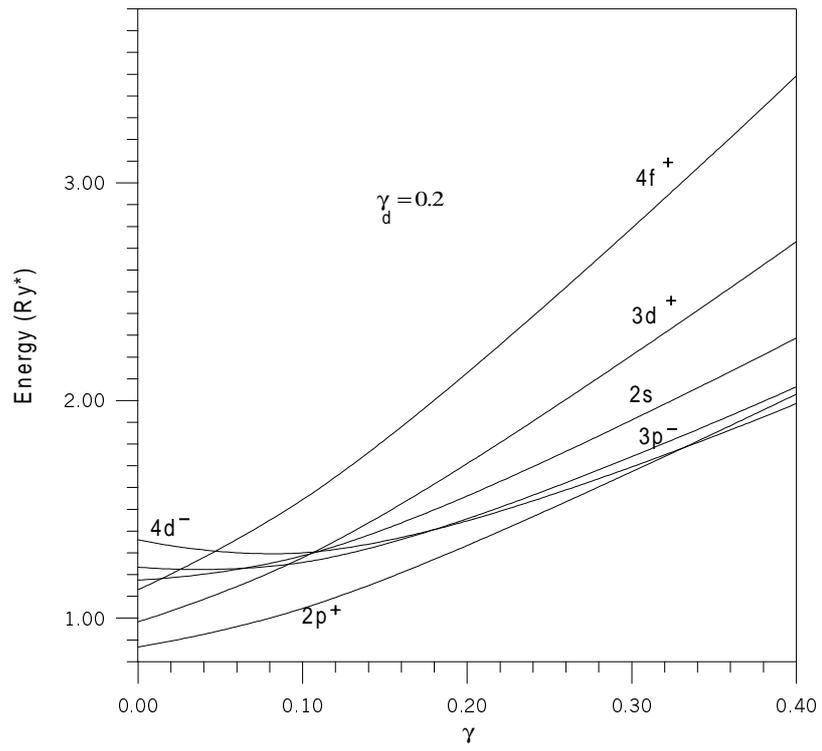
{\bf Fig.4:} Same as figure 2 for 3s, 3p$^+$, 4p$^-$, 5d$^-$,
            and 6f$^-$ states.\\

{\bf Fig.5:} Electron - electron interaction energies $E_{ee}(k,m)$
            versus $\Gamma$ for the states reported in tables 2 and 3.\\

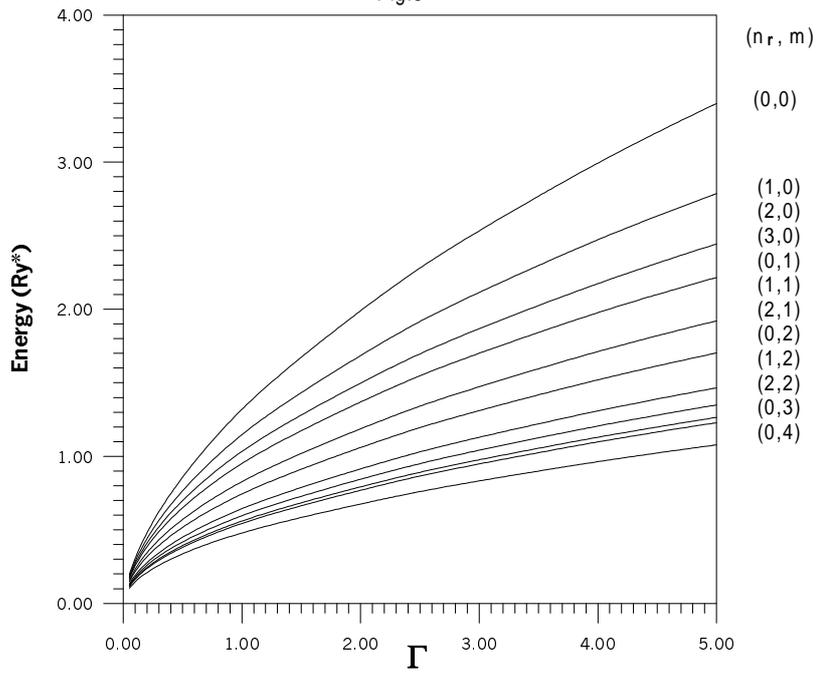
{\bf Fig.6:} $E(k,m;K,M)$ versus $\gamma$ for two electrons in QDs
            with $\gamma_d=0.2$ excluding the e$^-$ - e$^-$ interaction.\\

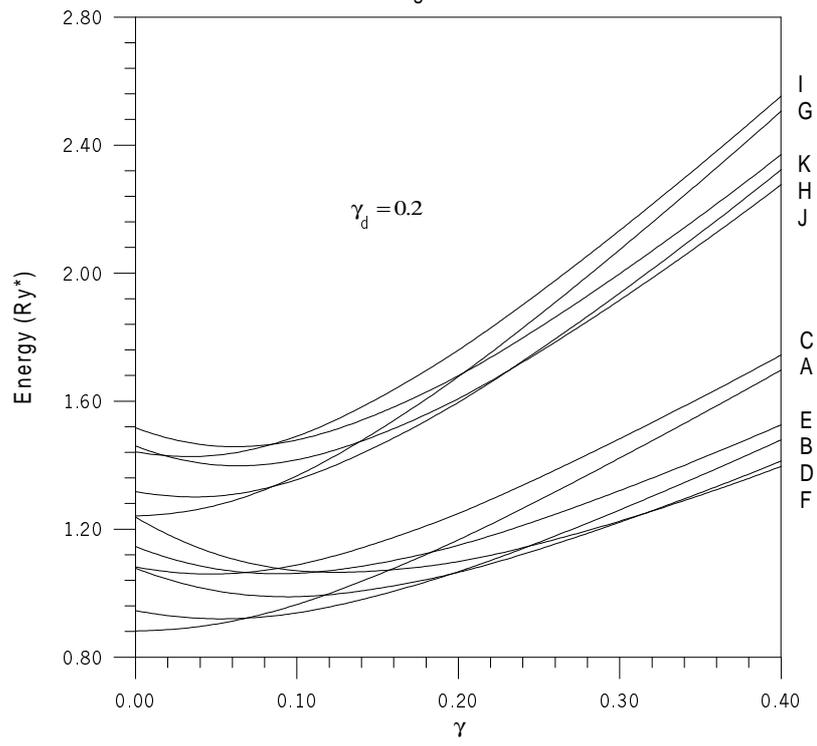
{\bf Fig.7:} Same as figure 6 including the e$^-$ - e$^-$ interaction.

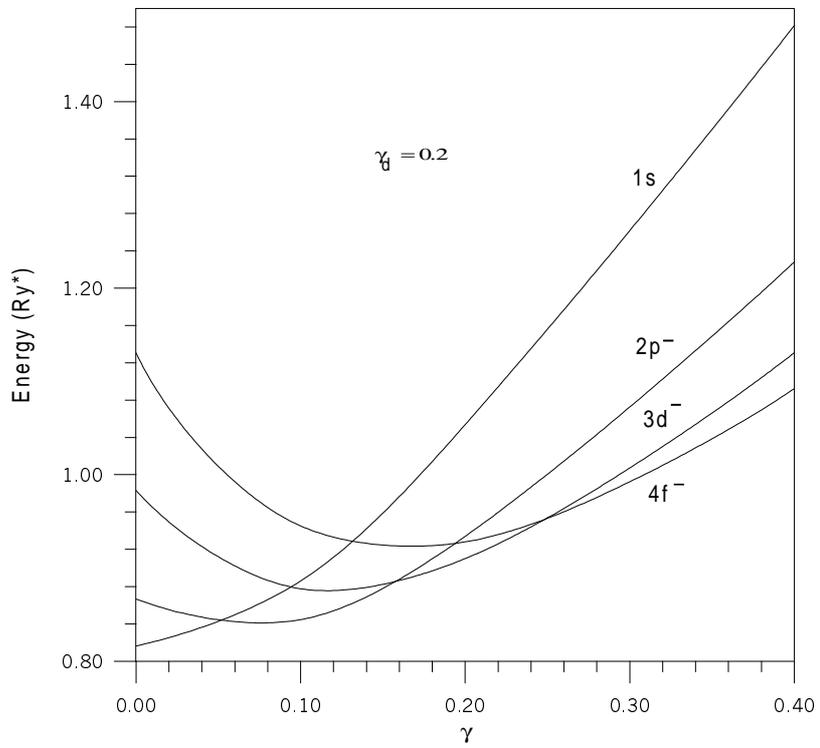
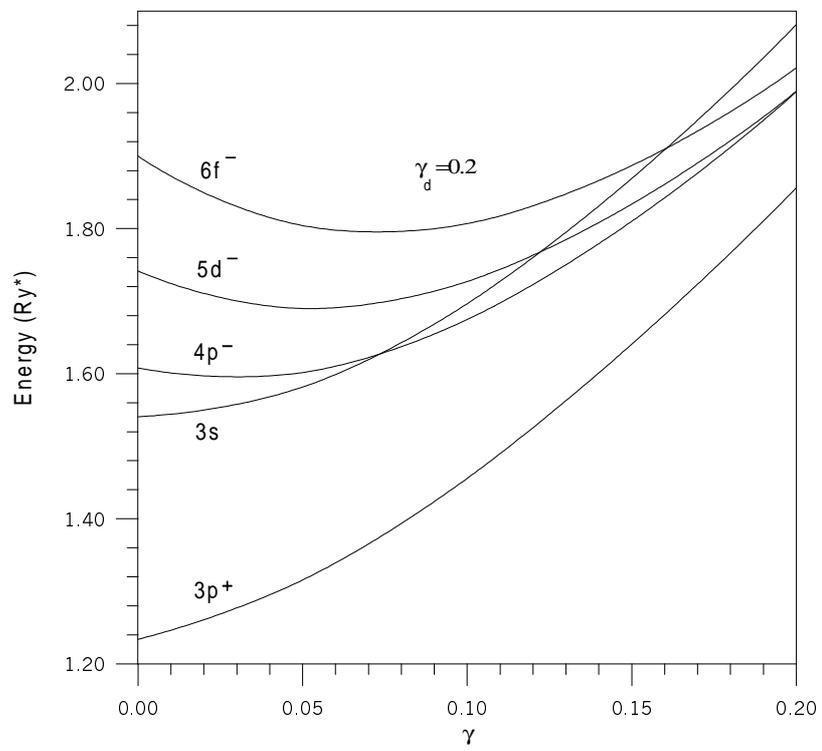
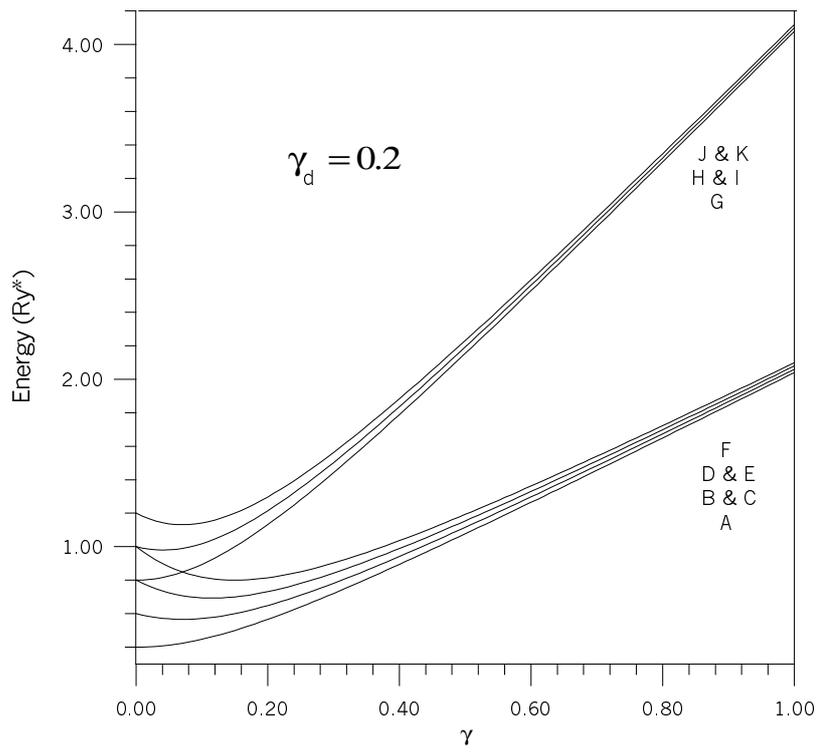
\end{document}